\newread\testifexists
\def\GetIfExists #1 {\immediate\openin\testifexists=#1
	\ifeof\testifexists\immediate\closein\testifexists\else
        \immediate\closein\testifexists\input #1\fi}
\def\epsffile#1{Figure: #1} 	

\immediate\openin\testifexists=e
	\ifeof\testifexists\immediate\closein\testifexists\else
        \immediate\closein\testifexists\input e\fipsf 
  
\magnification= \magstep1	
\tolerance=1600 
\parskip=5pt 
\baselineskip= 5 true mm \mathsurround=1pt
\hsize=5.3in  
\vsize=7.2in  
\font\smallrm=cmr8

\font\medrm=cmr9  
\font\medit=cmti9

\font\bigbf=cmbx12
 	\def\Bbb#1{\setbox0=\hbox{$\tt #1$}  \copy0\kern-\wd0\kern .1em\copy0} 
	\immediate\openin\testifexists=a
	\ifeof\testifexists\immediate\closein\testifexists\else
        \immediate\closein\testifexists\input a\fimssym.def 
\def\secbreak{\vskip10pt plus .5in \penalty-100\vskip 0pt plus -.4in} 
\def\biggerskip{\vskip6mm plus 2mm}
\def\hugeskip{\vskip9mm plus 3mm}
\def\Narrower{\par\narrower\noindent}	
\def\Endnarrower{\par\leftskip=0pt \rightskip=0pt} 
\def\br{\hfil\break}	\def\ra{\rightarrow}		
\def\a{\alpha}          \def\b{\beta}   \def\g{\gamma}  \def\G{\Gamma}
\def\d{\delta}          \def\D{\Delta}  \def\e{\varepsilon}
              \def\l{\lambda}         \def\L{\Lambda} 
\def\m{\mu}             \def\f{\phi}                
\def\n{\nu}             \def\j{\psi}    
\def\r{\varrho}         \def\s{\sigma}  
\def\t{\tau}                  
                     
\def\w{\omega}  \def\W{\Omega}                          

 \def\LL{{\cal L}} \def\OO{{\cal O}}  
\def\DD{{\cal D}}

\def\cl{\centerline}    
\def\ni{\noindent}      \def\pa{\partial}       \def\dd{{\rm d}}

\def\fn#1{\ifcase\noteno\def\fnchr{*}\or\def\fnchr{\dagger}\or\def
	\fnchr{\ddagger}\or\def\fnchr{\rm\S}\or\def\fnchr{\|}\fi
	\footnote{$^{\fnchr}$} 
	{\scrunch#1\toe}\ifnum\noteno>3 \global\advance\noteno by -5 \fi
	\global\advance\noteno by 1}
 	\def\scrunch{\baselineskip=11 pt \medrm}
 	\def\toe{\vphantom{$p_\big($}}
	\newcount\noteno

\def\fract#1#2{{\textstyle{#1\over#2}}}
\def\ffract#1#2{\raise .35 em\hbox{$\scriptstyle#1$}\kern-.25em/
	\kern-.2em\lower .22 em \hbox{$\scriptstyle#2$}}

\def\half{\fract12} \def\quart{\fract14}

\def\part#1#2{{\partial#1\over\partial#2}} 
 \def\ref#1{${\vphantom{)}}^#1$}

\def\bbf#1{\setbox0=\hbox{$#1$} \kern-.025em\copy0\kern-\wd0
        \kern.05em\copy0\kern-\wd0 \kern-.025em\raise.0433em\box0}              

 \def\bal{$\bullet$} 
\def\ref#1{${\,}^{\hbox{\smallrm #1}}$}

\def\Gbar{\raise.13em\hbox{--}\kern-.35em G}
\def\lap{\setbox0=\hbox{$<$}\,\raise .25em\copy0\kern-\wd0\lower.25em\hbox{$\sim$}\,}
\def\glt{\setbox0=\hbox{$>$}\,\raise .25em\copy0\kern-\wd0\lower.25em\hbox{$<$}\,}
\def\gap{\setbox0=\hbox{$>$}\,\raise .25em\copy0\kern-\wd0\lower.25em\hbox{$\sim$}\,}
   \def\newsect#1{\secbreak\noindent{\bf #1}\medskip}

  \rightline{SPIN-1998/18}
 \rightline{hep-th/9812203}
\medskip
\cl{\bf The glorious days of physics}\medskip
\cl{\bigbf RENORMALIZATION OF GAUGE THEORIES.}\bigskip
\cl{lecture notes Erice, August/September 1998.}

\hugeskip

\cl{\bf Gerard 't Hooft }
\biggerskip
\cl{Institute for Theoretical Physics}
\cl{University of Utrecht, Princetonplein 5}
\cl{3584 CC Utrecht, the Netherlands}
\smallskip
\cl{and}
\smallskip
\cl{Spinoza Institute}
\cl{Postbox 80.195}
\cl{3508 TD Utrecht, the Netherlands}
\smallskip\cl{e-mail: \tt g.thooft@phys.uu.nl}
\cl{internet: \tt http://www.phys.uu.nl/\~{}thooft/	}
\hugeskip
\ni{\bf Abstract:}\Narrower
   This is an account of the author's recollections of the
   turbulent days preceding the establishment of the Standard
   Model as an accurate description of all known elementary	particles
   and forces.
\Endnarrower
\bigskip\newsect{1. PREHISTORY ($\approx 1947 - 1969$): THE 
DEMISE OF QUANTUM FIELD THEORY.}

After a decade of great excitement due to the replacement of 
divergent integrals with the experimentally measured values
of the physical quantities (mass and charge), new and serious
troubles emerged that appeared to impede any further effective
understanding of the renormalization problem. 
In the '60s, two prototypes of
renormalizable quantum field theories were known to exist:
\item{1.} Quantum Electrodynamics (QED), a realistic model describing 
charged fermions interacting with the electromagnetic field, and
\item{2.} $\l\f^4$-theory, a theory in which scalar particles
interact among one another. In contrast to the first theory, this
model was not expected to apply to any of the elementary particles
known at the time.

\ni The general consensus was that {\it the real world is not described by
a renormalizable quantum field theory.}\ref{1, 2} With the benefit of hindsight, 
we can now identify what the reasons were for this misunderstanding.

In 1953, Peterman and Stueckelberg\ref3 noted an important aspect of
renormalized amplitudes. A 3-vertex, for instance, can be described 
as\fn{The notation here is different from the presentation in the lecture.
In these notes, the words `bare' and `renormalized' are used more carefully.}\br
\cl{\epsffile{glor1.ps}} or
$$\G\quad=\quad g^{\rm ren}\quad+\quad (g)^3\int(\cdots)\quad-\quad
\D g\quad .\eqno(1.1)$$
Thus, the full amplitude is built up from a lower order vertex, $g^{\rm ren}$,
a loop correction,
and a counter term $\D g$, to absorb the apparent infinities. It is clear that
the splitting between $g^{\rm ren}$ and $\D g$ is arbitrary, and the full
amplitude should not depend on that. It should only depend on the
`bare' coupling $g^{\rm bare}=g^{\rm ren}-\D g$. However, whenever we truncate the perturbative
expansion, the coupling constants used inside the multiloop diagrams are the renormalized
constants $g^{\rm ren}$. Therefore, in practice, some spurious dependence on the
splitting procedure remains. This should disappear when we add everything up, to all 
orders in perturbation theory.

The independence of the full amplitude on the subtraction procedure was
interpreted by Peterman and	Stueckelberg as a symmetry of the theory, and
it was called the renormalization group. The transformation is
$$\bigg\{\matrix{g^{\rm ren}&\ \ra\ &g^{\rm ren}+\e\ ;\cr
\D g&\ra&\D g\ -\,\e\ .\cr}\eqno(1.2)$$

This may appear to be a big symmetry, but its actual usefulness is limited to only
one -- albeit extremely important -- case.
It turns out that only {\it scale transformations\/} should be associated
with renormalization group transformations. This is a one-dimensional
subgroup of the renormalization group, and it is all that is still in use today.

In 1954, M.~Gell-Mann and F.~Low\ref4 observed that under a scale transformation for
the variable energy scale $\m$, the renormalization group transformation of
the fine structure constant $\a$ can be computed, and they found
$${\m\dd\over\dd\m}\a=\OO(\a^2)>0\,.\eqno(1.3)$$
In perturbation expansion, the function at the r.h.s. of Eq. (1.3) is a
Taylor series in $\a$, and it begins with a coefficient times $\a^2$.

In Moscow, L.~Landau expected this function to be ever-increasing, and hence
$\a(\m)$ should be an increasing function of $\m$, first increasing only very
slowly, because $\a(1\, {\rm MeV})$ is small, but eventually, its increase
should be explosive, and, even if the series in (1.3) should terminate
at $\a^2$, the function $\a(\m)$ should have a singularity at some finite
value of $\m$. This singularity is called the Landau pole, and it appears
to be a physically unacceptable feature. This was the reason why Landau,
and with him a large group of researchers, dismissed renormalized quantum
field theory as being mathematically flawed.

Gell-Mann and Low, on the other hand, speculated that the function at the r.h.s. of
Eq. (1.3) could have a zero, and in this case, the running coupling parameter
$\a(\m)$ should terminate at a certain value, being the bare coupling constant
of the theory. To compute this bare coupling constant, however, one would
have to go beyond the perturbation expansion, and it was not known how to
do this. So, although Gell-Mann and Low did not dismiss the theory, it
was clear that they demanded mathematical techniques that did not exist.
Consequently, not only in Eastern Europe, but also in the West, many physicists
believed that quantum field theory was based on mathematically very shaky
procedures.

All this hinged on the general belief that positivity of the renormalization
group function, later called beta-function, was inevitable. This belief was
based on the so-called K\"allen-Lehmann representation\ref5 of the propagator:
$$D(k^2)=\int{\r(m^2)\dd m^2\over k^2+m^2-i\e}\ ;\qquad	\r(m^2)>0\ .\eqno(1.4)$$
The function $\r(m^2)$ is positive as it gets its contributions from all
virtual states into which a particle can decay. The fact that the relation between
$\r$ and $\b$ is not so straightforward was apparently overlooked.
Renormalizable quantum field theory was regarded as a toy, and some researchers
claimed that the apparent numerical successes of quantum electrodynamics were
nothing more than accidental.\ref2
\newsect{2. INTERESTING MODELS.}
Yet, several of such deceptive toys continued to emerge as amusing models.
The most prominent one was proposed by C.N.~Yang and R.~Mills\ref6 in 1954. The fundamental
Lagrangian,
$$\LL^{\rm YM}=-\quart G_{\m\n}G_{\m\n}-\bar{\j}(\g D+m)\j\,,\eqno(2.1)$$
was charmingly simple and displayed a formidable symmetry: local gauge invariance.
Of course, it could not be used to describe the real world (it seemed), because
it required the presence of massless interacting vector particles that are different
from ordinary photons, more like electrically charged photons, and such particles
do not appear to exist. The closest things in the real world resembling them were
the $\r$ meson -- but this was more likely just an excited state of	hadronic matter
that happened to have spin one -- and the hypothetical intermediate force agent of the 
weak interaction, $W^\pm$, which could well be a vector particle. These however, have mass, and
no gauge-invariant term in the Lagrangian could generate such a mass.

Yet, this model continued to inspire many researchers. First there was R.~Feynman and Gell-Mann,
when they proposed a particular form for the fundamental Fermi interaction Lagrangian of the
weak force (for the record, this expression had been arrived at earlier by R.E.~Marshak
and E.C.G.~Sudarshan)\ref7:
$$\LL^{\rm weak}=G_W\bar{\j}\g_\m(1+\g_5)\j\ \bar{\j}\g_m(1+\g_5)\j\,,
\eqno(2.2)$$ where $G_W$ is an interaction constant.
It was not difficult to see that such an interaction could result from the
virtual creation and annihilation of heavy vector bosons, $W^\pm$.

Richard Feynman was again inspired by the Yang-Mills theory, when he was
investigating the mysteries of quantizing the gravitational force.\ref8 In gravity, the
relevant invariance group is that of the diffeomorphisms:
$$\f(x)\ra \f'(x)=\f(x+u(x))\,,\eqno(2.3)$$ which is local and non-Abelian, and thus
comparable to local gauge invariance in Yang-Mills theory:
$$\j(x)\ra\j'(x)=\W(x)\j(x)\,.\eqno(2.4)$$ It had been Gell-Mann who suggested
to Feynman to investigate the Yang-Mills theory instead of gravity, because it 
is simpler, while what had bugged Feynman was the non-Abelian nature of the
symmetry, a feature shared by the Yang-Mills system. Feynman discovered that, in
order to restore unitarity in this theory, spurious components must be added to
the Feynman rules. He called these ``ghosts". He could not go beyond one loop. A few years later, B.~DeWitt
formulated Feynman rules for higher loops.\ref9

While Feynman and DeWitt investigated massless gauge theories, S.~Glashow\ref{10} in 1961
added a mass term in order to obtain a decent looking Lagrangian for the weak
force: $$\LL=\LL^{\rm YM}-\half M^2 A_\m^2\,.\eqno(2.5)$$
This theory seemed to describe quite well the weak force, explaining its vector nature,
and also the apparent universality of the weak force, all particles subject to the weak
force having a universal coupling to the vector boson, as if there existed
a conserved Yang-Mills charge.

In 1964, P.~Higgs\ref{11} showed that an earlier theorem of J.~Goldstone\ref{12} does {\it not\/}
apply to local gauge theories. Goldstone had shown that whenever a continuous symmetry
is spontaneously broken by the vacuum state of a model, there must exist a massless,
spinless particle. Higgs showed that, if the symmetry is a {\it local\/} symmetry,
then Goldstone's particle is replaced by one that does have mass. It is now called
the `Higgs particle', but, even though Higgs avoided to use the words `field theory',
his paper called little attention at the time.

Shortly after, F.~Englert and R.~Brout\ref{13} derived that, if a local symmetry is spontaneously
broken, then not only Goldstone's particle, but also the gauge vector particle
develops a mass term. This is what is now called the `Higgs mechanism'.	All of this
happened when renormalized field theories were not {\it en vogue}, so the authors
used abstract mathematical arguments while avoiding specific models	such as the
Yang-Mills theory.

Abdus Salam\ref{14} made a strong case for Yang-Mills models with Higgs mechanism to be
used as weak interaction models. Independently, S.~Weinberg\ref{15}, in 1967, wrote down
the first detailed model for a combined electromagnetic and weak force with
a local $SU(2)\times U(1)$ symmetry, broken by the Higgs mechanism. 

There were two problems, however.  One was the renormalizability. Although these 
theories {\it seemed\/} to be renormalizable, the actual way to establish
computational rules could not be grasped. A second problem was, that although
Weinberg's theory appeared to work well for the leptons, the weak interactions
for the hadronic particles did not agree with experiment. The theory suggested
`strangeness-changing neutral current interactions', and these were ruled out
by observations.

The first problem was attacked by Veltman.\ref{16} Unimpressed by the arguments	of Higgs,
Englert and Brout, he started off from Glashow's Lagrangian  (2.5).	He found that,
up to one-loop diagrams, judicious manipulations of Feynman's ghosts would
suppress all infinities, and so he was encouraged to prove renormalizability up
to all orders. His computer algorithms in 1968 presented no clue. He all but proved
that none of these theories can be renormalized beyond one loop.

Another renormalizable model, to be applied to hadronic particles, was proposed
by Gell-Mann and M.~L\'evy\ref{17} in 1960. Their elementary fields were an isospin $\half$
nucleon field, $N=\pmatrix{p\cr n}$,  an isospin one pseudoscalar field for the
pions, $\vec\pi=\pmatrix{\pi^+\cr\pi^0\cr\pi^-}$, and a new scalar field $\s$.
This model, called the sigma model, had as its Lagrangian
$$\eqalign{\LL\,&=\,-\half(\pa\s^2+\pa{\vec\pi}^2)-V(\s^2+{\vec\pi}^2)-\bar N\g\pa N\cr
&-\,g\bar N(\s+i\vec\pi\cdot\vec\t\,\g_5)N\ +c\,\s\ .}\eqno(2.6)$$ 
Here, the components of $\vec\t$ are the isospin Pauli matrices, and $V$ stands
for a quadratic polynomial of its argument, $\s^2+\vec{\pi}^2$. Using the chiral
projection operators $P^\pm=\half(1\pm\g_5)$, one could see that this model has
a (global) chiral $SU(2)\times SU(2)\times U(1)_{\rm baryon}$ symmetry, broken only
by the last term, linear in $\s$. In this model, the symmetry can be spontaneously
broken down into $SU(2)\times U(1)$, if the potential $V$ has its minimum when its
argument does not vanish. The pions are then Goldstone's bosons,
their mass being proportional to the coefficient $c$ in (2.6). This model reflects
nicely the symmetry patterns observed in Nature. Already at that time, it was realized 
that these features could be explained in a quark theory. Writing
$$\vec\pi=i\bar q\vec\t\g_5 q\ ;\qquad \s=\bar q\,q\ ,\eqno(2.7)$$
the term $c\,\s$ corresponds to a quark mass term, which violates the quark's
chiral symmetry.

Sigma models are still often studied, but their origin in Gell-Mann and L\'evy's sigma
model is often forgotten. The physical features of this model strongly depend on
the symmetry of the vacuum state. If the pseudoscalar self interaction $V$ has its
minimum at the origin, the chiral symmetry is explicit, and the nucleons should 
be massless. Excited states of the nucleons should come in `parity doublets':
pairs of fermionic particles whose members have opposite parity. The pions and the
sigma all have the same, non vanishing mass. This is called the Wigner mode.

If the minimum of $V$ is away from the origin, we have spontaneous symmetry breaking,
the pions become massless but the nucleons have mass. Parity doublets are no longer
mass-degenerate. The sigma has mass also. The pion can only have mass if the symmetry
is broken explicitly, having $c\ne 0$. This we refer to as the Nambu-Goldstone mode.\ref{12}

The question was raised whether these distinctions survive renormalization, and
this was studied by B.W.~Lee\ref{18}, J.-L.~Gervais\ref{19} and K.~Symanzik.\ref{20} A Summer Institute
at Carg\`ese, 1970, was devoted to these studies. The outcome was that this model
is renormalizable, and its chiral symmetry properties are preserved after renormalization.
However, according to the measurements, the constant $g$, coupling the
pion field to the nucleons, is very large, and hence perturbative expansions are not
very fruitful for calculating the physical properties of nucleons and pions.
The sigma had to be so unstable that it could never be detected by experiments.
Attempts were made to improve these perturbative techniques using Pad\'e approximants and the
like.

Nowadays we can easily observe that the Yang-Mills theory, the theorems of Higgs,
Englert and Brout, and the sigma model were
among the most important achievements of the '50s and the '60's. However, most
physicists in those days would not have agreed. Numerous other findings were thought to
be much more important. As at prehistoric times, it may have seemed that the dinosaurs
were much more powerful and promising creatures than a few species of tiny, inconspicuous
little animals with fur rather than scales, whose only accomplishment was that 
they had developed a new way to reproduce and
nurture their young. Yet, it would be these early mammals that were decisive for
later epochs in evolution. In a quite similar fashion, Yang-Mills theories, quantum
gravity research and the sigma model were insignificant little animals with fur compared to the many
dinosaurs that were around: we had numerous strong interaction models, current algebras,\fn{Some 
of my friends feel that they are offended here: current algebra is not truely
extinct. Yet it did not evolve in the way expected at the time.}\ 
axiomatic approaches, duality and analyticity, which were attracting far more attention.
Most of these activities have now disappeared. 
\newsect{3. TAMING THE INFINITIES.}
In 1970, I understood that the Glashow model, Eq.~(2.5), which was also used by Feynman\ref8 and by
Veltman\ref{16}, had serious difficulties at the far ultraviolet, and that in this respect the
Higgs models were much to be preferred. Veltman, then my thesis advisor, did not 
agree. Finally, we could agree about the following program for my research:
\item{1).} Exactly how should one renormalize the amplitudes for the pure, massless
Yang-Mills system? What exactly are the calculational rules? Of course, this would
be done within the realm of perturbation theory.
\item{2).} And then how can one reconcile the procedure found with mass terms? I knew
that this was going to be the Higgs theory.

\ni The first step was not going to be easy. How does one subtract the infinities, and 
how does one compute the higher order corrections, such that local gauge invariance 
remains intact?	How is local gauge-invariance represented anyway, in a theory where
the gauge must be fixed by a gauge condition? We wanted the resulting theory to be
as precisely defined as quantum electrodynamics, $\l\,\f^4$ theory and the $\s$-model.

A beautiful study had been made of the functional integrals for a gauge theory by L.D.~Faddeev
and V.N.~Popov. Their functional integrals read as
$$\int\DD A_\m\ \D\ \exp \,i\int\LL^{\rm YM}\dd^4x\ \cdot\ \prod_x\d\big(\pa_\m A_\m
(x)\big)\ , \eqno(3.1)$$
where the Dirac delta in the last term is a gauge-fixing term, and the term $\D$ stands
for a determinant that is crucial for this expression to become independent of the gauge fixing. It
is a Jacobian, and the formal rules for its calculation could be given.
They were not unlike the rules for Feynman's ghost particle. The propagator for the
vector particle would have to be the transverse one:
$${\d_{\m\n}-k_\m k_\n/k^2\over k^2-i\e}\ .\eqno(3.2)$$

But there was also a paper by S. Mandelstam\ref{22}, who had derived a Feynman propagator
$${\d_{\m\n}\over k^2-i\e}\,.\eqno(3.3)$$ 
And Feynman had argued that the massless theory should be
seen as the limit of a massive theory, whose propagator is
$${\d_{\m\n}+k_\m k_\n/M^2\over k^2-i\e}\,,\eqno (3.4)$$
He claimed that, in the limit $M\ra 0$, the $k_\m k_\n$ term can be removed, and then it is
replaced by the effects of a ghost. His ghost rules did not exactly coincide with
those of Mandelstam and Faddeev and Popov.\fn{The reasons for this difference was already understood by 
Faddeev and Slavnov\ref{23}, as I learned later.}

I noted\ref{24} that the Faddeev-Popov determinant can be rewritten in terms of a Gaussian
functional integral, so that it can be seen to correspond to the contribution of a ghost:
$$\D={\rm det}(M)\ ;\qquad C\cdot({\rm det}\,M)^{-1}=\int\DD\f\DD\f^*\ e^{-\f^*M\f}\,.
\eqno(3.5)$$
The $-1$ in this equation explains the anomalous minus signs associated to ghosts.
Next, one could generalize the Faddeev-Popov procedure, replacing the gauge-fixing
term by $\d\left(\pa_\m A_\m-f(x)\right)$, after which one could integrate over $f(x)$
with an arbitrary Gauss term:
$$\int\DD A\DD\f\DD\f^*\DD f\ \D\ \exp\,i\int\left(\LL^{\rm YM}-\f^* M\f-\half\a f^2
\right)\prod_x\d\left(\pa_\m A_\m-f(x)\right)\ ,\eqno(3.6)$$
which produces a quite convenient extra term $-\half\a(\pa_\m A_\m)^2$ in the Lagrangian.
This turns the propagator into
$${\d_{\m\n}-\l k_\m k_\n/k^2\over k^2-i\e}\,,\eqno(3.7)$$
with arbitrary $\l$. So, we could see that Mandelstam's rules should give the same
amplitudes as Faddeev and Popov's.

In quantum electrodynamics, gauge invariance is reflected in Ward identities. What were
the exact Ward identities in Yang-Mills theories? We identified the combinatorial 
properties of Feynman diagrams that lie at the basis of Ward identities of 
electrodynamics, and these could be generalized to understand the corresponding
identities in gauge theories. This was quite a tour de force\ref{24}, as we could not identify
the global symmetry that could be used to derive these\fn{The relevant global symmetry
is now known to be BRS symmetry\ref{28}, but this has anticommuting generators, which we did
not know how to use in those times. In this symmetry, also, the Jacobi identity is essential.}. 
An essential ingredient was that the
gauge group must be complete, i.e., the generators of the group must obey the
Jacobi identity. Diagrammatically, we have for instance:\br 
\cl{\epsffile{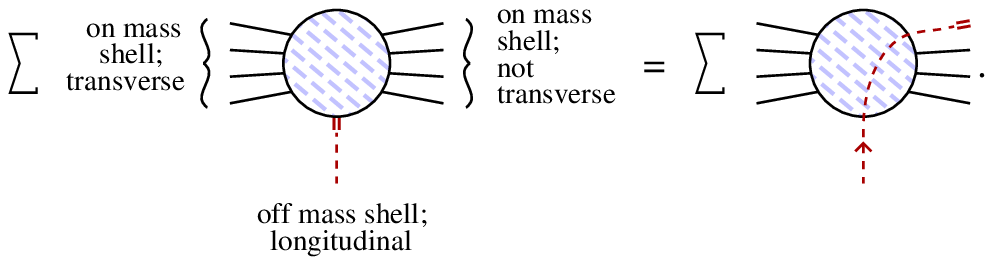}}
These are the identities one needs, to prove unitarity of the subtraction procedures.
The intermediate states in the identity
$$S\ S^\dagger\ =\ {\Bbb I}\,,\eqno(3.8)$$ 
cancel due to ghost contributions of the form (among others)\br 
\cl{\epsffile{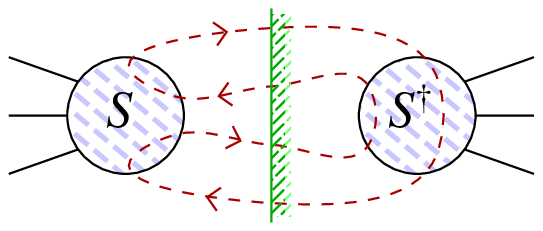}}

We then established that these identities are precisely what is needed to determine
all renormalization counter terms completely and unambiguously.	There was a problem,
however. In chiral theories, it was known that {\it anomalies\/} could occur. Different
identities may require different choices for the finite parts of the counter terms,
such that there is a clash. Can such clashes occur in gauge theories? After an extensive
search, a regulator technique was found that is gauge-invariant, and therefore it
obeys the identities automatically. We simply introduced a fifth dimension\ref{24} in Minkowski space,
and chose the value of the momentum in that direction, inside any loop diagram, 
to be a fixed number, $\L$. This
worked, but only for one-loop diagrams. We still had to prove uniqueness for the
multi-loop case. The fifth-dimension method we had at that time would later turn out
to be the precursor to a more general dimensional regularization procedure which
would solve the multi-loop problem.

I was most eager to proceed to the next step: add the mass terms. This was easy	now.
Absolute gauge-invariance was absolutely essential, so it was forbidden to add the mass
terms as in the Glashow model. But the Higgs mechanism was just fine. It was now
easy to generalize our procedures to include the Higgs mechanism. Again, unitarity and
all other required properties could be established perturbatively.\ref{25} 

By this time, two papers appeared: J.~Taylor\ref{26} and A.A.~Slavnov\ref{27} noticed independently
that my identities could be extended to off-mass shell amplitudes. In my earlier
approach, I had decided to avoid making such a step since such identities would
necessitate the introduction of new counter terms, and so they would cause
complications. However, it was these off-shell identities that could be 
interpreted in terms of BRS symmetry\ref{28}, and so it happened that the pivotal
relations among amplitudes needed to establish renormalizability are now
referred to in the literature as ``Slavnov-Taylor identities".

We still had to address the question whether the finite parts of the counter terms
may suffer any anomaly or not at the multi-loop level. Adding a 6th or 7th dimension
to Minkowski space did not lead to unambiguous answers\fn{A book by a Russian
author appeared in which such multi-dimensional regulators were employed. 
Its proofs are incorrect.}. Finally, with M.~Veltman\ref{29}, a correct procedure was
launched, now called ``dimensional regularization and renormalization". The
theory must be taken at $n=4-\e$ dimensions. Physically, non-integral dimensions
would be meaningless, but in {\it perturbation expansion}, the amplitudes
{\it generated by Feynman diagrams\/} can be uniquely defined. The logarithmic
divergences disappear immediately; the linear and quadratic divergences
persist, but if $\e\ne 0$ they can be subtracted {\it unambiguously\/} by
partial integration techniques. One is then left with poles of finite order
in the complex $\e$ plane. These poles can be removed from the physical amplitudes
by means of gauge-invariant counter terms (field renormalizations may require non-gauge
invariant subtractions, but fields are not directly observable in this scheme).

Dimensional regularization was not only good for formal proofs, but it also turned out 
to be a very practical tool for calculating renormalized higher loop diagrams. For the first
time, we had a theory in which higher order corrections to weak interaction effects were
finite and could be computed. However, it was not yet
clear which of the possible gauge models most accurately described the observed
interactions. A solution to the hadron problem had already been proposed earlier
by Glashow, J.~Iliopoulos and L.~Maiani\ref{30}, who introduced  a new quark
species named `charm'. Their mechanism could explain the absence of strangeness-changing
neutral current events, but strangeness-conserving neutral currents should persist.
Events due to these currents were detected in beautiful experiments, both in the
hadronic sector and in the leptonic sector.
\newsect{4. SCALING.}
Because of the phenomenological successes of gauge theories, many physicists 
abandoned their previous objections against renormalizable field theories.
Yet the problems with scaling appeared to persist.
In 1970, independently, C.G.~Callan\ref{31} and K.~Symanzik\ref{31}	wrote down the equations
among amplitudes following from scaling modified by renormalization group effects.
They used functions	of the coupling parameters $g$ that were called 
$\a(g)$, $\b(g)$, $\g(g)$, \dots, of which the function $\b(g)$ played the same
role as the function mentioned in Sect.~1. As they concentrated on the familiar
prototypes of all renormalizable theories, QED and $\l\f^4$, they expected 
this function always to be positive. Yet, experiments on inelastic scattering
at very high energies suggested nearly naive scaling properties there, as if
$\b(g)\lap 0$, and the coupling constants $g$ themselves were very small
(Bjorken scaling\ref{36}). The sentiment against quantum field theories induced D.~Gross in 1971 to
conjecture that {\it no quantum field theory will be able to describe Bjorken
scaling}.\ref{33}

What about the Yang-Mills theories? Unusual signs had been obtained for related
effects in vector particle theories at various instances. In 1964, a calculation
was performed by V.S. Vanyashin and M.V. Terentev\ref{34}, who found that the charge
renormalization of charged vector bosons is negative. This result was considered
to be absurd, and attributed to the non-renormalizability of this theory. The
charge renormalization in Yang-Mills theory was calculated correctly by
Khriplovich\ref{35} in 1969, again resulting in the unusual sign, but the connection with
asymptotic freedom was not made, and his remarkable work was ignored.

When I studied the renormalizability of gauge theories, I was of course interested
in scaling, and I started calculations in 1971. The calculation is tricky, because
of the following. The self-energy corrections, including the ghost contribution,
take the form $C\,(k_\m k_\n-k^2\d_{\m\n})$, suggesting a counter term of the form 
$$C (\pa_\m A_\n-\pa_\n A_\m)^2\,,\eqno(4.1)$$  The 3-vertex 
corrections	give a scaling correction to this vertex of the form
$$C' f^{abc}\pa_\m A^a_\n[A_\m^b A_\n^c]\,.\eqno(4.2)$$
 However, the coefficients $C$ and $C'$ do not match to form a gauge-invariant
 combination of the form $F^a_{\m\n}F^a_{\m\n}$. This was because the field
 renormalization is not gauge-independent. But the correct scaling behaviour
 can be deduced from these calculations. In any case, the sign was 
 unmistakable.\fn{I made a clear statement about this sign in my 1971
 paper on the renormalization procedure for Higgs theories\ref{25}. In this
paper, I motivated my work on these theories as follows (referring to the
previous paper): {\medit Thus, our prescription for the renormalization 
precedure is consistent, so the ultraviolet problem for mass-less Yang-Mills 
fields has been solved. A much more complicated problem is formed by the
infrared divergences of the system. [...] The disaster is such that the 
perturbation expansion breaks down in the infrared region, so we have no
rigorous field theory to describe what happens.} "Field theory" always meant
perturbative field theory in those days.}

At that time, I found it hard to publish anything without Veltman's approval,
and in this topic, he was simply disinterested. When I mentioned my
ideas about pure gauge fields coupling to quarks, he clearly stated that, if I had
any theory for the forces between quarks, I would have to explain why
physical quarks are not seen. This, I did not know at the time, but I
planned to try to find out. Before having good ideas about confinement, no publication
would be worth-while\dots

With so
many experts active in this field, I thought that the scaling behaviour of Yang-Mills theories
would probably be known to them anyway. I could not understand why quantum field theories
were categorically dismissed whenever Bjorken scaling was discussed. 
But in a small meeting at Marseille, June 1972,
I discussed with Symanzik his work on $\l\f^4$ theory with negative $\l$. He hoped to
be able to explain Bjorken scaling\ref{36} this way. When I explained to him that
gauge theories were a much better choice, because of what I had found out
about their scaling, he clearly expressed disbelief. But after he had
presented his work in the meeting\ref{37}, he gave me the opportunity to announce
publicly the following statement: {\it If you scale a gauge theory containing
vector, spinor and scalar particles, then the gauge coupling constant scales
according to}
$${\m\dd\over\dd\m} g^2={g^4\over 8\pi^2}\left(-\fract{11}3 C_1+\fract23 C_2
+\fract16 C_3\right)\,,\eqno(4.3)$$
where $C_1$ is a positive Casimir number associated to the gauge group,
$C_2$ a positive coefficient pertinent to the fermions, and $C_3$, also positive, belongs to
the scalars. Symanzik encouraged me to 
publish this result quickly, since it would be novel indeed. I now regret not to have
followed his sensible advice. Instead, I continued my work with Veltman
on the divergences in quantum gravity.\ref{38}

It was clear that the scaling expressed in Eq.~(4.2) was closely related to
the counter term required in dimensional renormalization. But this
relation is far from trivial. It was worked out in detail in 1973,
and in addition, a calculational procedure was presented, using background fields,
such that gauge-invariance can be exploited for the direct derivation
of gauge-invariant counter terms. This greatly simplified the derivation
of Eq.~(4.3). By that time, the papers by H.D.~Politzer\ref{39}, D.~Gross and F.~Wilczek\ref{40} had
appeared.
\newsect{5. CONFINEMENT AND MONOPOLES.}
So, finally, pure gauge theories for quarks were taken seriously. The roots
for the theory now called quantum chromodynamics, date from 1964, when
O.W.~Greenberg\ref{41} made an attack on the spin-statistics connection for
quarks inside hadrons. In 1965, M.~Han and 	Y.~Nambu\ref{42} proposed a model
that would now be called a Higgs theory. $SU(3)^{\rm color}\times U(1)$ would be
broken spontaneously into a subgroup $U(1)^{\rm EM}$ in such a way that the 
physical electric charges of the quarks would turn up as integral multiples
of $e$. This is actually quite close to modern QCD\fn{See my other lecture
at this School.}. W.A.~Bardeen, H.~Fritzsch, and M.~Gell-Mann\ref{43} described pure $SU(3)$
in 1972, not as a field theory but as a current algebra. Gross, Wilczek, Fritzsch, Gell-Mann
and H.~Leutwyler\ref{44} then proposed a pure $SU(3)$ Yang-Mills system in 1973,
but they stressed that there were two main problems: the apparent absolute
confinement of quarks, and the axial $U(1)$ problem. Related ideas were
put forward by H.~Lipkin\ref{45} in 1968 and 1973.
 
The first indication concerning the nature of quark confining forces
came from an analysis of dual resonance models. G.~Veneziano\ref{46} had proposed
a formula for elastic meson-meson scattering that fits well with phenomenological 
observations concerning hadron scattering. When a generalization of this expression
for inelastic scattering involving higher particle multiplicities was constructed
by Z.~Koba and H.~Nielsen\ref{47}, a physical interpretation was discovered by 
D.~Fairlie and H.B.~Nielsen\ref{48}, 
Nambu\ref{49}, T.~Goto\ref{50} and L.~Susskind\ref{51}: these amplitudes 
describe strings with quarks at
their end-points. Then, Nielsen and P.~Olesen\ref{52}, and B.~Zumino\ref{53}
discovered a model that leads to such strings from an ordinary field theory,
the Abelian Higgs model. Just as inside a super conductor, this model allows
for magnetic fields to penetrate the vacuum, but only if they are squeezed
to form tight vortex lines.

A drawback of this model, however, was that it required quarks to carry
strong magnetic charges, very unlike their physical characteristics in QCD.
I was interested in  understanding whether such vortex-like structures could arise
naturally in QCD. I decided to do a little exercise: what are the characteristics
of a magnetic vortex in a non-Abelian Higgs theory? 
This exercise led to a surprise: in a non-Abelian theory, magnetic vortices
may be unstable. If two vortices are arranged parallel one to another, they
might annihilate. This was counter to our intuition; if vortices carry
magnetic flux, then this annihilation should only be allowed if single
magnetic north or south charges can be created in the theory.

I was led to the conclusion that if a local non-Abelian, compact gauge symmetry
is broken into $U(1)$ by the Higgs mechanism, then, with respect to the
$U(1)$ photons,  localised field solutions should exist that, in all respects,
behave as isolated magnetic charges. Knowing that magnetic monopoles had
been speculated about for a long time in quantum field theories, I realized that
this was a major discovery. After publishing it\ref{54}, I was made aware of a
similar discovery made by A.~Polyakov\ref{55} in the Soviet Union, independently
but somewhat later.

The first indication of the real confinement mechanism came from an
idea by K.~Wilson.\ref{56} He considered Yang-Mills theory on a lattice, and then
studied the $1/g^2$ expansion. At all orders, one finds that quarks are
linked together by strings. These strings represent the {\it electric\/}
flux, rather than the magnetic flux as in Nielsen and Olesen.

Apparently, what is needed to understand confinement, is an interchange
electric--magnetic, which in many respects is equivalent to a weak coupling -- 
strong coupling interchange. This led to the idea that we should search for
a new Higgs effect, where the condensed particles are not electrically
but magnetically charged. Since magnetic monopoles occur naturally in
a gauge theory, a crude description of this mechanism could be given.\ref{57}

Then, V.~Gribov\ref{58}	came with a quite different observation: he noted
an ambiguity when gauge-fixing is performed in a non-perturbative
calculation. At first, my impression was that his suggestion that this
may be related to the confinement problem, was far-fetched. But later,
I realized that he had a point. In any case, a most accurate approach to
the non-perturbative quantization of the theory required an unambiguous
gauge-fixing, and I discovered that such a requirement would suggest a
two-step procedure to understand confinement: first, one fixes the gauge
partially, but unambiguously and without ghosts, such that only an
Abelian gauge-invariance is left\ref{59}, and then one quantizes the remaining
system as if it were QED. The nice thing about this procedure is that,
after the first step, one does not exactly obtain an Abelian gauge theory,
but one which also contains magnetic monopoles. The latter could easily
condense, at which point the electric field lines would form vortices
that are precisely of the kind needed to confine quarks.
This is how we now understand quark confinement to work. The vindication
of these ideas came in the late '70s from computer simulations in lattice models.

Wilson had pioneered lattice gauge theory. However, he treated the system as
a statistical theory at a coupling strength close to its critical point. It took some time
before it was realized that, probably, the effective coupling in QCD is not
at all close to its critical point. In that case, simple computer simulations
could be sufficient to establish the essential dynamical features.
It was M.~Creutz, L.~Jacobs and C.~Rebbi\ref{60} who surprised the community with
their first successful computer simulations on quite small lattices.

Presently, lattice simulations have reached impressive accuracies,
so that their results can be checked with experimental data on hadronic resonances.\ref{61}
\newsect{6. THE $U(1)$ PROBLEM.}

Permanent quark confinement was not the only puzzle that had to be solved
before the more tenacious critics would accept QCD as being a realistic
theory.  There has been another cause for concern.  This was the so-called
$\eta$-$\eta'$ problem.  The problem was evident already to the earliest
authors, Fritzsch, Gell-Mann and Leutwyler\ref{44}.  The QCD Lagrangian
correctly reflects all known symmetries of the strong interactions:
\par\bal \  baryon conservation, symmetry group $U(1)$; 
\par\bal \  isospin, symmetry group $SU(2)$, broken by electromagnetism, 
and also a small mass difference between the up and the down quark.  
\par\bal \  an $SU(3)$ symmetry, broken more strongly by the strange quark  mass.
\par\bal \  spontaneously broken chiral $SU(2)\times SU(2)$, also explicitly
broken by the mass terms of the up and the down quark.  The explicit chiral
symmetry breaking is reflected in the small but non-vanishing value of the
pion mass-squared, relative to the mass-squared of other hadrons. 
\par\bal \  spontaneously broken chiral $SU(3)\times SU(3)$, explicitly broken 
down to $SU(2)\times SU(2)$ by the strange quark mass, which leads to a 
relatively light kaon.

\ni Furthermore, the hadron spectrum is reasonably well explained by
extending these symmetries to an approximate $U(6)$ symmetry spanned by the
$d$, $u$ and $s$ quark states, each of which may have its spin up or down.

The problem, however, is that the QCD Lagrangian also shows a symmetry that
is known to be badly broken in the observed hadron spectrum:  instead of
chiral $SU(2)\times SU(2)$ symmetry (together with a single $U(1)$ for baryon
number conservation), the Lagrangian has a chiral $U(2)\times U(2)$, and
consequently, it suggests the presence of an additional axial $U(1)$ current
that should be approximately conserved.  This should imply that a fourth
pseudoscalar particle (the quantum numbers are those of the $\eta$ particle)
should exist that is as light as the three pions.  Yet, the $\eta$ is
considerably heavier:  549 MeV, instead of 135 MeV.  For similar reasons, one
should expect a ninth pseudoscalar boson that is as light as
the kaons.  The only candidate for this would be one of the heavy mesons, 
originally called the $X^0$~meson. When it was discovered that this meson
can decay into two photons\ref{62}, and therefore had to be a 
pseudoscalar, it was renamed as  $\eta'$. However, this $\eta'$ was 
known to be as heavy as 958 MeV, and it appeared to be impossible
to accommodate for this large chiral $U(1)$ symmetry breaking by modifying the
Lagrangian.  Also, the decay ratios -- especially the radiative decays --
 of $\eta$ and $\eta'$ appeared to be
anomalous, in the sense that they did not obey theorems from chiral algebra.
\ref{63}
\midinsert\cl{\epsffile{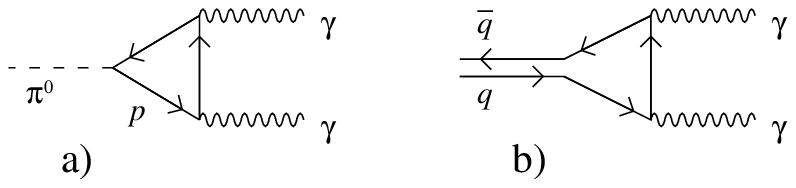}}
\cl{\scrunch Fig.~1. The U(1) anomaly, a) in terms of virtual protons; b) in terms
of virtual quarks.}
\endinsert

A possible cure for this disease was also recognised:  the Adler-Bell-Jackiw
anomaly. Its history goes back to J.~Steinberger's calculation of the 
$\pi^0$ decay into two photons. This process can be attributed to the creation
of a virtual $p\,\bar p$ pair (Fig. 1a), or in modern terminology, the creation
of virtual $q\,\bar q$ pairs (Fig.~1b), but it appears to violate the
chiral $SU(2)\times SU(2)$ symmetry, and would have been forbidden in the
$\s$-model, if there hadn't been an anomaly. The axial vector current is not conserved\ref{64};
its continuity equation is corrected by quantum effects. In QCD, there is just such a
correction: $$\pa_\m J^A_\m={g^2\over16\pi^2} \e_{\m\n\a\b}\,{\rm
Tr}\,G_{\m\n}G_{\a\b}\,,$$ where $J^A_\m$ is the axial vector current,
$G_{\m\n}$ the Yang Mills gluon field, and $g$ the strong coupling constant.
So, again, the axial current is not conserved.  Then, what is the problem?

The problem was that, in turn, the r.h.s.  of this anomaly equation can also
be written as a divergence:  $$\e_{\m\n\a\b}\,{\rm
Tr}\,G_{\m\n}G_{\a\b}=\pa_\m K_\m \,,$$ where $K_\m$ is the Chern-Simons
current.  $K_\m$ is not gauge-invariant, but it appeared that the latter
equation would be sufficient to render the $\eta$ particle as light as the
pions.  Why is the $\eta$ so heavy?

There were other, related problems with the $\eta$ and $\eta'$ particles:
their mixing.  Whereas the direct experimental determination of the
$\w$-$\phi$ mixing\ref{65} allowed to conclude that in the octet of
vector mesons, $\w$ and $\f$ mix in accordance to their quark contents:
$$\w=\fract{1}{\sqrt2}(u\bar u+d\bar d)\ ,\qquad\f=s\bar s\ ,$$ the $\eta$
particle is strongly mixed with the strange quarks, and $\eta'$ is nearly an
$SU(3)$ singlet:  
$$\eta\approx\fract12(u\bar u+d\bar d-\sqrt2\, s\bar s)\
,\qquad \eta'\approx\fract12(u\bar u+d\bar d+\sqrt2\, s\bar s)\ .$$ Whence
this strong mixing?

Several authors came with possible cures.  J.~Kogut and Susskind\ref{66}
suggested that the resolution came from the quark confinement mechanism, and
proposed a subtle procedure involving double poles in the gluon propagator.
S.~Weinberg\ref{67} also suggested that, somehow, the would-be Goldstone boson
should be considered as a ghost, cancelling other ghosts with opposite
metrics.  My own attitude was that, since $K_\m$ is not
gauge-invariant, it does not obey the boundary conditions required to allow
one to do partial integrations, so that it was illegal to deduce the
presence of a light pseudoscalar. Still, this argument was not yet quite satisfactory.

Just as it was the case for the confinement problem, the resolution to this
$U(1)$ problem was to be found in the very special topological structure of
the non-Abelian forces.  In 1975, a topologically non-trivial field
configuration in four-dimensional Euclidean space was described by four
Russians, A.A.~Belavin, A.M.~Polyakow, A.S.~Schwarz and
Yu.S.~Tyupkin\ref{68}. It was a localized configuration that featured a
fixed value for the integral $$\int\dd^4x\,\e_{\m\n\a\b}\,{\rm
Tr}\,G_{\m\n}G_{\a\b}={32\pi^2\over g^2}\,.$$ The importance of this finding
is that, since the solution is localized, it obeys all physically reasonable
boundary conditions, and yet, the integral does not vanish.  Therefore, the
Chern-Simons current does not vanish at infinity.  Certainly, this thing had
to play a role in the violation of the $U(1)$ symmetry.

This field configuration, localized in space as well as in time, was to be
called ``instanton'' later\ref{69}. Instantons are sinks and sources for the
chiral current.  This should mean that chirally charged fermions are created
and destroyed by instantons.  How does this mechanism work?

Earlier, R.~Jackiw and C.~Rebbi\ref{70} had found that the Dirac equation
for fermions near a magnetic monopole shows anomalous zero-energy solutions,
which implies that magnetic monopoles can be given fractional chiral quantum
numbers.  We now discovered that also near an instanton, the Dirac equation
shows special solutions, which are fermionic modes with vanishing
action\ref{69}. This means that the contribution of fermions to the
vacuum-to-vacuum amplitude turns this amplitude to zero!  Only amplitudes in
which the instanton creates or destroys chiral fermions are unequal to zero.
The physical interpretation of this was elaborated further by Russian
investigators, by Jackiw, Nohl and Rebbi\ref{71}, and by C.~Callan,
R.~Dashen and D.~Gross\ref{72}: instantons are tunnelling events.  Gauge
field configurations tunnel into other configurations connected to the
previous ones by topologically non-trivial gauge transformations.  During
this tunnelling process, one of the energy levels produced by the Dirac
equation switches the sign of its energy.  Thus, chiral fermions can pop up
from the Dirac sea, or disappear into it.  In a properly renormalized
theory, the number of states in the Dirac sea is precisely defined, and
adding or subtracting one state could imply the creation or destruction of
an antiparticle.  This is why the original Adler-Bell-Jackiw anomaly\ref{64} was
first found to be the result of carefully renormalizing the theory.

With these findings, effective field theories could be written down in such
a way that the contributions from instantons could be taken into account as
extra terms in the Lagrangian.  These terms aptly produce the required mass
terms for the $\eta$ and the $\eta'$, although it should be admitted that
quantitative agreement is difficult to come by; the calculations are
exceptionally complex and involve the multiplication of many large and small
numerical coefficients together.  But it is generally agreed upon
(with a few exceptions) that the $\eta$ and $\eta'$ particles behave as they
should in QCD\fn{There were protests.  R.~Crewther continued to
disagree\ref{73}. It was hard to convince him\ref{74}.}
\newsect{7. FROM MODEL TO THEORY.}
With the explanation of the confinement mechanism, confinement being
demonstrated as a generic feature of pure gauge theories on a lattice, and
the resolution of the eta mass problem and the eta mixing problem, the last
obstacles against the acceptance of QCD were removed. A beautiful observation by
Sterman and Weinberg\ref{75} was that, at very high energies, the QCD constituent
particles -- quarks and gluons -- can in practice be observed by registering
the showers of particles in their wakes: jets. When, for instance, quarks annihilate into
gluons, one sees a jet where each gluon was supposed to be. Measurements show that
these jets can be accurately predicted using this interpretation. It is even
possible to measure accurately the decrease of the QCD coupling strength.

But a single experimental discovery put an end to any lingering doubts.
This was the advent of the $J/\j$ particle. It was quickly recognised that this
particle had to be viewed as a charm-anticharm bound state, and for theoreticians,
it confirmed two fundamental hypotheses at once: first, it confirmed the GIM
mechanism. Although more indirect evidence for charm had already been reported,
we now had a genuine charm factory, and masses and couplings of the charmed
quark could now be measured. Secondly, it confirmed that the forces between
quarks decrease dramatically with decreasing distances. At small distance
scales, one can employ perturbative QCD to compute many details of this
beautiful system. We now gained real confidence that we were understanding
the basic forces of Nature. We could combine the electro-weak model with
QCD to obtain an accurate description of Nature, called the Standard Model. 
Indeed, up till now, our models had been regarded as severely simplificated caricatures
of the real world. Now, for the first time, we could treat the combination we
had obtained, as a {\it theory}, not just a model. The Standard Model is
very accurately correct. It should be called the Standard Theory.
\newsect{REFERENCES.}

\item{1.} See the arguments raised in: R.P.~Feynman, in {\it The Quantum 
Theory of Fields -- The 12th
Solvay Conference}, 1961 (Interscience, New York).  
\item{2.} T.Y.~Cao and S.S.~Schweber, {\it The Conceptual Foundations and
Philosophical Aspects of Renormalization Theory}, {\it Synthese  \bf 97} (1993)
33, Kluwer Academic Publishers, The Netherlands.
\item{3.} E.C.G.~Stueckelberg and A.~Peterman, {\it Helv.~Phys.~Acta  
\bf  26} (1953) 499; N.N.~Bogoliubov and D.V.~Shirkov, {\it 
Introduction  to  the  theory  of quantized fields }
(Interscience, New York, 1959);
A.~Peterman, {\it Phys.~Reports  \bf  53c} (1979) 157.
\item{4.} M.~Gell-Mann and F.~Low, {\it Phys.~Rev.  \bf  95} (1954) 1300.
\item{5.} G.~K\"allen, {\it Helv.~Phys.~Acta \bf 23} (1950) 201, {\it ibid.}
{\bf 25} (1952) 417; H.~Lehmann,
{\it Nuovo Cimento  \bf 11} (1954) 342.
\item{6.} C.N.~Yang and R.L.~Mills, {\it Phys.~Rev.} {\bf  96} (1954) 191, {\it
see also:\/} R.~Shaw,  Cambridge   Ph.D.~Thesis, unpublished. 
\item{7.} R.P.~Feynman and M.~Gell-Mann, {\it Phys.~Rev.} {\bf  109} (1958) 193; see the
earlier work by E.C.G.~Sudarshan and R.E.~Marshak, Proc.~Padua-Venice Conf. on {\it Mesons 
     and Recently discovered Particles}, 1957, p.~V - 14,  reprinted in: 
     P.K.~Kabir, {\it Development of Weak Interaction  Theory},  Gordon  and 
     Breach, 1963, p.~118; 
E.C.G.~Sudarshan and R.E.~Marshak, {\it Phys.~Rev.} {\bf  109} (1958) 1860.
\item{8.} R.P.~Feynman, {\it Acta Phys.~Polonica \bf 24} (1963) 697.
\item{9.} B.S.~DeWitt, {\it Phys.~Rev.}~Lett. {\bf 12} (1964) 742, and    
  {\it Phys.~Rev.} {\bf  160} (1967) 1113; {\it ibid. \bf 162} (1967) 1195, 1239.
\item{10.} S.L.~Glashow, {\it Nucl.~Phys.} {\bf 22} (1961) 579.                                    
\item{11.} P.W.~Higgs, {\it Phys.~Lett. \bf 12 }(1964) 132; {\it Phys.~Rev.}~Lett. {\bf   13 } (1964)  508; 
     {\it Phys.~Rev.} {\bf  145} (1966) 1156.
\item{12.} J.~Goldstone, {\it Nuovo Cim.~\bf 19} (1961) 15; Y.~Nambu and G.~Jona-Lasinio,
{\it Phys.~Rev. \bf 122} (1961) 345.     
 \item{13.} F.~Englert and R.~Brout, {\it Phys.~Rev.~Lett.} {\bf 13} (1964) 321.    
\item{14.} A.~Salam and J.C.~Ward, {\it Nuovo Cim.}{\bf  19} (1961) 165
A.~Salam and J.C.~Ward, {\it Phys.~Lett.} {\bf 13} (1964) 168
A.~Salam, Nobel Symposium 1968, ed.~N.~Svartholm     
\item{15.} S.~Weinberg, {\it Phys.~Rev.~Lett. \bf 19} (1967) 1264.
\item{16.} M.~Veltman, {\it Physica \bf 29} (1963) 186,
{\it  Nucl.~Phys. \bf B7} (1968) 637; J.~Reiff and  M.~Veltman, 
{\it Nucl.~Phys.~\bf B13} (1969) 545; M.~Veltman, {\it Nucl.~Phys. \bf B21} (1970) 288.  
\item{17.} M.~Gell-Mann and M.~L\'evy, {\it  Nuovo Cim.  \bf  16} (1960) 705.
\item{18.} B.W.~Lee, {\it Chiral Dynamics}, Gordon and Breach, New York (1972)
\item{19.} J.-L.~Gervais and B.W.~Lee, {\it Nucl.~Phys. \bf B12} (1969) 627;
	J.-L.~Gervais, Carg\`ese lectures, July 1970.
\item{20.} K.~Symanzik, Carg\`ese lectures, July 1970.
\item{21.} L.D.~Faddeev and V.N.~Popov, {\it Phys.~Lett. \bf 25B} (1967) 29;
L.D.~Faddeev, {\it Theor.~and Math.~Phys. \bf 1} (1969) 3 (in Russian),
 {\it Theor.~and Math.~Phys. \bf 1} (1969) 1 (Engl. transl).
\item{22.} S.~Mandelstam, {\it Phys.~Rev.} {\bf  175} (1968) 1580, 1604.
\item{23.} L.D.~Faddeev and A.A.~Slavnov, {\it Theor.~Math.~Phys. \bf 
3} (1970) 18 (English trans. on page 312).  
\item{24.} G.~'t Hooft, {\it Nucl.~Phys.} {\bf B33} (1971) 173.  
\item{25.} G.~'t Hooft, {\it Nucl.~Phys.} {\bf B35} (1971) 167.   
\item{26.} J.C.~Taylor, {\it Nucl.~Phys.~\bf B33} (1971) 436.
\item{27.}  A.~Slavnov, {\it Theor.~Math.~Phys. \bf 10} (1972) 153 (in Russian), 
{\it Theor.~Math.~Phys. \bf 10} (1972) 99 (Engl. Transl.) 31.  
\item{28.} C.~Becchi, A.~Rouet and R.~Stora, {\it Commun.~Math.~Phys. \bf 42} (1975) 127; 
     id., {\it Ann.~Phys.} (N.Y.) {\bf 98} (1976) 287; see also    
     I.V.~Tyutin, Lebedev Prepr. FIAN 39 (1975), unpubl.
\item{29.} G.~'t Hooft and M.~Veltman, {\it Nucl.~Phys.} {\bf B44} (1972) 189; {\it Nucl.~Phys. \bf B50} (1972) 318.
See also:     C.G.~Bollini and J.J.~Giambiagi,  {\it Phys.~Letters} {\bf 40B}  (1972) 
     566; J.F.~Ashmore, {\it Nouvo Cim. Letters \bf 4} (1972) 289.
\item{30.} S.L.~Glashow, J.~Iliopoulos and L.~Maiani, {\it Phys.~Rev.} {\bf  D2} (1970) 1285. 
\item{31.} C.G.~Callan, {\it Phys.~Rev.} {\bf  D2} (1970) 1541.
\item{32.} K.~Symanzik, Commun.~Math.~Phys. {\bf 16} (1970) 48;
	{\it ibid.} {\bf 18} (1970) 227.
\item{33.} D.J.~Gross, in {\it The Rise of the Standard Model},
Cambridge Univ.~Press (1997), p.~199
\item{34.} V.S.~Vanyashin, M.V.~Terentyev, {\it Zh.~Eksp.~Teor.~Phys.  \bf 48} (1965) 565 
[Sov. Phys. JETP 21 (1965)].
\item{35.} I.B.~Khriplovich, {\it Yad.~Fiz. \bf 10} (1969) 409  [{\it Sov.~J.~Nucl.~Phys. \bf 10} (1969)].
\item{36.}  J.D.~Bjorken, {\it Phys.~Rev. \bf 179} (1969) 1547;
 R.P.~Feynman, {\it Phys.~Rev.~Lett. \bf 23} (1969) 337. 
\item{37.} K.~Symanzik, in Proc.~Marseille Conf. 19-23 June 1972, ed. C.P.~Korthals 
     Altes;  id.,  {\it Nuovo Cim.~Lett.  \bf 6} (1973) 77.   
\item{38.} G.~'t Hooft and M.~Veltman, {\it Ann.~Inst.~Henri Poincar\'e,  \bf 20} (1974) 69.
\item{39.}	H.D.~Politzer, {\it Phys.~Rev.~Lett. \bf 30} (1973) 1346.
\item{40.} D.J.~Gross and F.~Wilczek, {\it Phys.~Rev.~Lett.  \bf30} (1973) 1343. 
\item{41.} O.W.~Greenberg, {\it Phys.~Rev.~Lett. \bf 13} (1964) 598.
\item{42.} M.Y.~Han and Y.~Nambu, {\it Phys.~Rev. \bf 139 B} (1965) 1006; 
	Y.~Nambu, in {\it Preludes in Theoretical Physics}, ed.~A.~de-Shalit et 
     al, North Holland Pub.~Comp., Amsterdam 1966, p.~133
\item{43.} W.A.~Bardeen, H.~Fritzsch and M.~Gell-Mann, in the Proceedings of the
	Frascati Meeting May 1972, {\it Scale and Conformal Symmetry
	in Hadron Physics}, ed. R.~Gatto, John Wiley \& Sons, p.~139. 
\item{44.} H.~Fritsch, M.~Gell-Mann and H.~Leutwyler, {\it Phys.~Lett.} {\bf 47B} (1973) 365.
\item{45.} H.J.~Lipkin, {\it Phys.~Lett.}~{\bf 45B} (1973) 267;  
See also: H.J.~Lipkin, in {\it Physique Nucl\'eaire},  Les-Houches  1968,  ed.
     C.~DeWitt and V.~Gillet, Gordon and  Breach,  N.Y.~1969,  p.~585; 
\item{46.} G.~Veneziano, {\it Nuovo Cimento \bf 57A} (1968) 190.
\item{47.} Z.~Koba and H.B.~Nielsen, {\it Nucl.~Phys.  \bf B10} (1969) 633,   
{\it ibid.  \bf B12} (1969) 517; {\bf B17} (1970) 206; {\it
Z.~Phys.  \bf 229} (1969) 243.
\item{48.} H.B.~Nielsen, "An Almost Physical Interpretation of the Integrand of the 
  n-point Veneziano model, {\it XV Int.  Conf.  on  High  Energy  Physics}, 
  Kiev, USSR, 1970; Nordita Report (1969), unpublished;
  D.B.~Fairlie and H.B.~Nielsen, {\it Nucl.~Phys. \bf B20} (1970) 637.
\item{49.}  Y.~Nambu, 
 Proc.~Int.~Conf. on {\it Symmetries and Quark models},
Wayne state Univ. (1969); Lectures at the Copenhagen Summer Symposium (1970).
\item{50.} T.~Goto, {\it Progr.~Theor.~Phys. \bf 46} (1971) 1560.
\item{51.} H.B.~Nielsen and L.~Susskind, CERN preprint TH 1230 (1970);
	L.~Susskind, {\it Nuovo Cimento \bf 69A} (1970) 457; {\it Phys.~Rev \bf 1} (1970) 1182.
\item{52.} H.B.~Nielsen and P.~Olesen, {\it Nucl.~Phys. \bf B61} (1973) 45.
\item{53.} B.~Zumino, in {\it Renormalization and Invariance in Quantum Field 
Theory},
NATO Adv.~Study Institute, Capri, 1973, Ed.~R.~Caianiello, Plenum (1974)
p.~367.
\item{54.}  G.~'t~Hooft, {\it Nucl.~Phys.  \bf B79} (1974) 276.
\item{55.} A.M.~Polyakov, {\it JETP Lett.  \bf 20} (1974) 194.
\item{56.} K.G.~Wilson, {\it Pys.~Rev. \bf D10} (1974) 2445.
\item{57.} G.~'t~Hooft, ``Gauge Theories  with  Unified  Weak,  Electromagnetic  and  
   Strong Interactions", in {\it E.P.S. Int. Conf.  on  High  Energy  Physics},  
   Palermo, 23-28 June 1975, Editrice Compositori, Bologna 1976, A.~Zichichi Ed.; 
 S.~Mandelstam, {\it Phys. Lett. \bf B53} (1975) 476; {\it Phys.  
	Reports \bf 23} (1978) 245. 
\item{58.} V.~Gribov, {\it Nucl.~Phys. \bf B139} (1978) 1.
\item{59.} G.~'t~Hooft, {\it Nucl.~Phys. \bf B190} (1981) 455.
\item{60.} M.~Creutz, L.~Jacobs and C.~Rebbi, {\it Phys.~Rev.~Lett. 
	\bf 42} (1979) 1390.
\item{61.} A.J.~van der Sijs, invited talk at the Int.~RCNP Workshop on Color 
Confinement and Hadrons -- Confinement 95, March 1995, RCNP Osaka, 
Japan (hep-th/9505019)
\item{62.} D.~Bollini {\it et al, Nuovo Cimento \bf 58A} (1968) 289.
\item{63.} A.~Zichichi, {\it Proceedings of the 16th International 
Conference on "High Energy Physics"\/}, Batavia, IL, USA, 6-13 Sept.~1972 
(NAL, Batavia, 1973), Vol {\bf 1}, 145. 
\item{64.} S.L.~Adler, {\it Phys.~Rev.}~{\bf  177} (1969) 2426; J.S.~Bell and R.~Jackiw, 
{\it Nuovo Cim.} {\bf  60A} (1969) 47 (zie ook bij Adler); S.L.~Adler and W.A.~Bardeen, 
{\it Phys.~Rev.}~{\bf  182} (1969) 1517;  W.A.~ Bardeen, 
     {\it Phys.~Rev.}~{\bf  184} (1969) 1848.
\item{65.} D.~Bollini {\it et al\/}, {\it Nuovo Cimento \bf 57A} (1968) 404.
\item{66.} J.~Kogut and L.~Susskind, {\it Phys.~Rev. \bf D9} (1974) 3501, {\it ibid.
\bf D10} (1974) 3468, {\it ibid. \bf D11} (1975) 3594.
\item{67.} S.~Weinberg, {\it Phys.~Rev \bf D11} (1975) 3583.
\item{68.} A.A.~Belavin, A.M.~Polyakov, A.S.~Schwartz and Y.S.~Tyupkin,  
	{\it Phys.~Lett. \bf 59 B} (1975) 85.
\item{69.} G.~'t~Hooft, {\it Phys.~Rev.~Lett.  \bf37} (1976) 8;
 {\it Phys.~Rev.  \bf  D14} (1976) 3432; Err.{\it  Phys.~Rev. \bf  D18} (1978) 2199.
\item{70.} R.~Jackiw and C.~Rebbi,  {\it Phys.~Rev.~\bf D13} (1976) 3398.
\item{71.} R.~Jackiw, C.~Nohl and C.~Rebbi, various Workshops on QCD and 
	solitons, June--Sept.~1977; R.~Jackiw and C.~Rebbi, 
	{\it Phys.~Rev.~Lett. \bf 37} (1976) 172.
\item{72.} C.G.~Callan, R.~Dashen and D.~Gross, 
		{\it Phys.~Lett. \bf 63B} (1976) 334.
\item{73.}  R.~Crewther, {\it Phys.~Lett. \bf 70 B} (1977) 359; {\it Riv.~Nuovo Cim.
 \bf 2} (1979) 63; R.~Crewther, in {\it Facts and Prospects of Gauge Theories}, 
 Schladming 1978, ed. P.~Urban (Springer-Verlag 1978),; 
 {\it Acta Phys.~Austriaca} Suppl {\bf XIX} (1978) 47.
\item{74.} G.~'t~Hooft, {\it Phys.~Repts.  \bf 142} (1986) 357.
\item{75.} G.~Sterman and S.~Weinberg, {\it Phys.~Rev.~Lett. \bf 39} (1977) 1436. 

\bye